\documentclass[pra,aps,showpacs]{revtex4}

\newcommand*{\Tr}{\mathop{\mathrm{Tr}}}

\newcommand*{\re}{\mathop{\mathrm{Re}}}

\begin{document}

\title{Influence of the detector's temperature on the quantum Zeno effect}

\author{Julius Ruseckas}
\email{ruseckas@itpa.lt}
\affiliation{Institute of Theoretical Physics and Astronomy,\\
 A. Go\v{s}tauto 12, 2600 Vilnius, Lithuania}
\date{\today{}}

\begin{abstract}
  In this paper we study the quantum Zeno effect using the irreversible model of
  the measurement. The detector is modeled as a harmonic oscillator interacting
  with the environment. The oscillator is subjected to the force, proportional
  to the energy of the measured system. We use the Lindblad-type master equation
  to model the interaction with the environment. The influence of the detector's
  temperature on the quantum Zeno effect is obtained. It is shown that the
  quantum Zeno effect becomes stronger (the jump probability decreases) when the
  detector's temperature increases.
\end{abstract}

\pacs{03.65.Xp, 03.65.Ta, 03.65.Yz}

\maketitle

\section{\label{sec:intro}Introduction}

The quantum Zeno effect is a consequence of the influence of the measurements on
the evolution of a quantum system. In quantum mechanics the short-time behavior
of the non-decay probability of unstable particle is not exponential but
quadratic \cite{Khalfin}. This deviation from the exponential decay has been
observed by Wilkinson \textit{et al.\/{}} \cite{GG}. In 1977, Misra and
Sudarshan \cite{Misra} showed that this behavior when combined with the quantum
theory of measurement, based on the assumption of the collapse of the wave
function, leaded to a very surprising conclusion: frequent observations slowed
down the decay. An unstable particle would never decay when continuously
observed. Misra and Sudarshan have called this effect the quantum Zeno paradox
or effect. Later it was realized that the repeated measurements could not only
slow the quantum dynamics, but the quantum process may be accelerated by
frequent measurements as well. This effect was called a quantum anti-Zeno
effect \cite{Kaulakys,Kofman,rus2}.

The quantum Zeno effect has been experimentally proved \cite{Itano} in a
repeatedly measured two-level system undergoing Rabi oscillations.  The
interruption of Rabi oscillations has been at the focus of interest
\cite{16,17,18,19,20,21,22,Kaulakys}. Recently, the quantum Zeno effect has been
considered for tunneling from a potential well into the continuum \cite{23}, as
well as for photoionization \cite{24}. The quantum anti-Zeno effect has been
obtained experimentally \cite{Fisher}.

In the analysis of the quantum Zeno effect the finite duration of the
measurement becomes important and, therefore, the projection postulate is not
sufficient to solve this problem. In Ref.~\cite{rus2} a simple model that allows
to take into account the finite duration and finite accuracy of the measurement
has been developed. However, this model does not take into account the
irreversibility of the measurement process.

The basic ideas of a quantum measurement process were theoretically expounded in
Refs.~\cite{Giuilini,Zurek1,Zurek2,Zurek3,Zurek4,Walls,Walls2} on the assumption
of environmentally induced decoherence or superselection.  In this paper we
extend the model, used in Ref.~\cite{rus2}, including the interaction of the
detector with the environment. Then it becomes possible to study the influence
of other parameters of the detector on the evolution of the measured system,
too. In this paper we analyze the influence of the detector's temperature on the
measured system.

To describe the decoherence and dissipation we use the the Lindblad-type master
equation. Semigroup theories pioneered by Lindblad \cite{Lindblad} demonstrated
that density-matrix positivity, translational invariance and approach to thermal
equilibrium cannot be satisfied simultaneously.  Under the assumption of
Markovian dynamics and initial decoupling of system and bath, the semigroup
approach adds dissipative dynamics to the quantum master equations by means of
the Lindblad dissipation operators.

Recently the semigroup formalism has attracted much attention.  Quantum
computing is one of the fields in which quantum dissipation finds the most
recent applications. In physical chemistry semigroup theories have been utilized
to model dynamics of ultrafast predissociation in a condensed-phase or cluster
environment \cite{Burghardt}, and electronic quenching due to the coupling of
the adsorbate negative ion in resonance to the metal electrons in the desorption
of neutral molecules on metal surfaces \cite{Saal}. In nuclear physics, the
semigroup formalism is applied to model giant resonances in the nuclear spectra
above the neutron emission threshold \cite{Stef}.

We proceed as follows: In Sec.~\ref{sec:model} we present the model of the
measurement. The method of the solution is presented in Sec.~\ref{sec:solution}.
The measurement of the unperturbed system is considered in
Sec.~\ref{sec:unpert-meas}.  In Sec.~\ref{sec:pert-meas} we derived a formula
for the probability of the jump into another level during the measurement of the
frequently measured perturbed system. Sec.~\ref{sec:concl} summarizes our
findings.

\section{\label{sec:model}Model of the measurement}

We consider a system that consists of two parts. The first part of the system
has the discrete energy spectrum. The Hamiltonian of this part is $\hat{H}_0$.
The other part of the system is represented by Hamiltonian $\hat{H}_1$.
Hamiltonian $\hat{H}_1$ commutes with $\hat{H}_0$. In a particular case the
second part can be absent and $\hat{H}_1$ can be zero. The operator $\hat{V}(t)$
causes the jumps between different energy levels of $\hat{H}_0$. Therefore, the
full Hamiltonian of the system is of the form
$\hat{H}_S=\hat{H}_0+\hat{H}_1+\hat{V}(t)$.  The example of such a system is an
atom with the Hamiltonian $\hat{H}_0$ interacting with the electromagnetic
field, represented by $\hat{H}_1$, while the interaction between the atom and
the field is $\hat{V}(t)$.

We will measure in which eigenstate of the Hamiltonian $\hat{H}_0$ the system
is. The measurement is performed by coupling the system with the detector. The
full Hamiltonian of the system and the detector equals to
\begin{equation}
\hat{H}=\hat{H}_S+\hat{H}_D+\hat{H}_I,
\label{eq:ham}
\end{equation}
where $\hat{H}_D$ is the Hamiltonian of the detector and $\hat{H}_I$ represents
the interaction between the detector and the measured system, described by the
Hamiltonian $\hat{H}_0$. As the detector we use a harmonic oscillator with the
Hamiltonian
\begin{equation}
\hat{H}_D=\hbar\Omega\left(\hat{b}^{\dag}\hat{b}+\frac{1}{2}\right),
\end{equation}
where $\hat{b}$ and $\hat{b}^{\dag}$ are the creation and anihillation
operators, respectively. We choose the interaction operator $\hat{H}_I$ in the
form
\begin{equation}
\hat{H}_I=\lambda\hat{q}\hat{H}_0,
\end{equation}
where $\hat{q}=\hat{b}^{\dag}+\hat{b}$ is the coordinate of the detector and the
parameter $\lambda$ describes the strength of the interaction. This system ---
detector interaction is similar to that considered by von Neumann \cite{vNeum}
and in Refs.~\cite{rus2,joos,caves,milb,gagen,rus1}.  In order to obtain a
sensible measurement, the parameter $\lambda$ must be large.

The measurement begins at time moment $t_0$. At the beginning of the interaction
with the detector, the detector's density matrix is $\hat{\rho}_D(t_0)$. The
detector initially is in the thermal equilibrium with the temperature $T$.
Therefore,
\begin{equation}
  \hat{\rho}_D(t_0)=\hat{\rho}_T=\exp\left(-\frac{\hbar\Omega\hat{n}}{k_BT}
  \right)\left(1-\exp\left(-\frac{\hbar\Omega}{k_BT}\right)\right),
\label{eq:rhoT}
\end{equation}
where $\hat{n}=\hat{b}^{\dag}\hat{b}$. The average excitation of the detector in
thermal equilibrium with the temperature $T$ is
\begin{equation}
\bar{n}(T)=\left(\exp\left(\frac{\hbar\Omega}{k_BT}\right)-1\right)^{-1}.
\end{equation}
The full density matrix of the system and detector is
$\hat{\rho}(t_0)=\hat{\rho}_S(t_0)\otimes\hat{\rho}_D(t_0)$ where
$\hat{\rho}_S(t_0)$ is the density matrix of the system.

The detector is interacting with the environment. The master equation for the
density matrix of the system and the detector in the Lindblad form is
(Ref.~\cite{Lindblad})
\begin{equation}
\label{eq:lindblad}
\frac{\partial\hat{\rho}(t)}{\partial t}=\frac{1}{i\hbar}
[\hat{H},\hat{\rho}(t)]+L_D[\hat{\rho}(t)],
\end{equation}
where
\begin{equation}
L_D[\hat{\rho}(t)]=\sum_{\mu}([\hat{V}_{\mu}\hat{\rho}(t),\hat{V}_{\mu}^{\dag}]
+[\hat{V}_{\mu},\hat{\rho}(t)\hat{V}^{\dag}_{\mu}]) ,
\label{eq:LD}
\end{equation}
and $\hat{V}_{\mu}$ are the Lindblad dissipation operators.  We use the equation
of a dissipative phase damped oscillator discussed in quantum optic
\cite{Giuilini}. The Lindblad dissipation operators are chosen as follows,
\begin{equation}
  \hat{V}_1=\sqrt{\frac{\gamma}{2}}\hat{a}^{\dag}\hat{a},
  \qquad\hat{V}_2=\sqrt{\frac{\gamma_{\uparrow}}{2}}\hat{a}^{\dag},
  \qquad\hat{V}_3=\sqrt{\frac{\gamma_{\downarrow}}{2}}\hat{a}.
\end{equation}
Then the equation (\ref{eq:lindblad}) for the density matrix becomes
\begin{eqnarray}
  \frac{\partial\hat{\rho}(t)}{\partial t} & = &\frac{1}{i\hbar}
  [\hat{H},\hat{\rho}(t)]+\frac{\gamma}{2}(2\hat{n}\hat{\rho}(t)\hat{n}-
  \hat{n}^2\hat{\rho}(t)-\hat{\rho}(t)\hat{n}^2)\nonumber\\
  &  & +\frac{\gamma_{\uparrow}}{2}(2\hat{a}^{\dag}\hat{\rho}(t)\hat{a}
  -(\hat{n}+1)\hat{\rho}(t)-\hat{\rho}(t)(\hat{n}+1))\nonumber\\
  &  & +\frac{\gamma_{\downarrow}}{2}(2\hat{a}\hat{\rho}(t)\hat{a}^{\dag}
  -\hat{n}\hat{\rho}(t)-\hat{\rho}(t)\hat{n}).
  \label{eq:lindbladtmp}
\end{eqnarray}
The approach to the thermal equilibrium is obtained when the parameters
$\gamma_{\uparrow}$ and $\gamma_{\downarrow}$ satisfy the condition
\cite{Rajagopal}
\begin{equation}
\gamma_{\uparrow}=\gamma_{\downarrow}\exp\left(-\frac{\hbar\Omega}{k_BT}\right).
\end{equation}

\section{\label{sec:solution}Solution of the master equation}

For the solution of the equation (\ref{eq:lindbladtmp}) we adopt the technique used
in Ref.~\cite{Zhao}. We introduce the quantum characteristic function
\cite{Gardiner}
\begin{equation}
\chi(\xi ,\xi^*)=\Tr\{\hat{\rho}e^{\xi\hat{b}^{\dag}}e^{-\xi^* \hat{b}}\}.
\end{equation}
The quantum characteristic function of the detector at the thermal equilibrium is
\begin{equation}
\chi_T(\xi ,\xi^*)=\exp(-\xi\xi^*\bar{n}(T)).
\end{equation}

We multiply the equation (\ref{eq:lindblad}) by $\exp(-\xi^*\hat{b})$ from the
left and by $\exp(\xi\hat{b}^{\dag})$ from the right and take the trace. When
the interaction between the measured system and the detector is absent (i.e.,
$\lambda =0$), we obtain the equation
\begin{eqnarray}
\frac{\partial}{\partial t}\chi(\xi ,\xi^*;t) & = & i\Omega\left(
\xi\frac{\partial}{\partial\xi}\chi -\xi^*\frac{\partial}{\partial\xi^*}
\chi\right)\nonumber\\
 &  & +\frac{\gamma}{2}\left(2\xi^*\xi
\frac{\partial^2}{\partial\xi\partial\xi^*}\chi -\xi^{*2}
\frac{\partial^2}{\partial\xi^{*2}}\chi -\xi^2\frac{\partial^2}{\partial\xi^2}
\chi -\xi^*\frac{\partial}{\partial\xi^*}\chi -\xi\frac{\partial}{\partial\xi}
\chi\right)\nonumber\\
 &  & +\frac{\gamma_{\uparrow}}{2}\left(\xi\frac{\partial}{\partial\xi}\chi
 +\xi^*\frac{\partial}{\partial\xi^*}\chi -2\xi\xi^*\chi\right)\nonumber\\
 &  & -\frac{\gamma_{\downarrow}}{2}\left(\xi\frac{\partial}{\partial\xi}
\chi +\xi^*\frac{\partial}{\partial\xi^*}\chi\right) .
\label{eq:nomeas} 
\end{eqnarray}
We will search the solution of Eq.~(\ref{eq:nomeas}) in the form
\begin{equation}
\label{eq:chi}
\chi(\xi ,\xi^*)=\exp\left(\sum_{j,k}C_{j,k}(t)\xi^j(-\xi^*)^k\right),
\end{equation}
where $C_{j,k}$ are the coefficients to be determined. Substituting
Eq.~(\ref{eq:chi}) into Eq.~(\ref{eq:nomeas}) we obtain the set of equations for
the coefficients $C_{j,k}$
\begin{eqnarray}
\frac{\partial C_{j,k}(t)}{\partial t} & = & i\Omega(j-k)C_{j,k}(t)
-\frac{\gamma}{2}(j-k)^2C_{j,k}(t)\nonumber\\
 &  & +\frac{1}{2}(\gamma_{\uparrow}-\gamma_{\downarrow})(j+k)C_{j,k}(t)
+\gamma_{\uparrow}\delta_{j,1}\delta_{k,1} .
\label{eq:coef} 
\end{eqnarray}
The solution of Eq.~(\ref{eq:coef}) is
\begin{eqnarray}
C_{1,1}(t) & = & C_{1,1}(0)e^{-(\gamma_{\downarrow}-\gamma_{\uparrow})t}
+\bar{n}(T)\left(1-e^{-(\gamma_{\downarrow}-\gamma_{\uparrow})t}\right),\\
C_{j,k}(t) & = & C_{j,k}(0)e^{i\Omega(j-k)t}e^{-\frac{\gamma}{2}(j-k)^2t
-\frac{1}{2}(\gamma_{\downarrow}-\gamma_{\uparrow})(j+k)t},
\quad j\neq 1,k\neq 1
\end{eqnarray}
From the solution we see that the function $\chi$ approaches the function at the
equilibrium $\chi_T$ as the time $t$ grows. The detector's density matrix
$\hat{\rho_D}$, correspondingly, tends to the $\hat{\rho_T}$.

\section{\label{sec:unpert-meas}Measurement of the unperturbed system}

At first, we will consider the case when the perturbation is absent, i.e.,
$\hat{V}(t)=0$. Since the Hamiltonian of the measured system does not depend on
$t$ we will omit the parameter $t_0$ in this section. We can choose the basis
$|n\alpha\rangle$ common for the operators $\hat{H}_0$ and $\hat{H}_1$,
\begin{eqnarray}
\hat{H}_0|n\alpha\rangle  & = & E_n|n\alpha\rangle ,\\
\hat{H}_1|n\alpha\rangle  & = & E_1(n,\alpha)|n\alpha\rangle ,
\end{eqnarray}
where $n$ numbers the eigenvalues of the Hamiltonian $\hat{H}_0$ and $\alpha$
represents the remaining quantum numbers.

We introduce the density matrix $\hat{\rho}_{m,n}=\sum_{\alpha}\langle m\alpha
|\hat{\rho}|n\alpha\rangle$ and the characteristic function
\begin{equation}
\chi_{m,n}(\xi ,\xi^* ;t)=\Tr\{\hat{\rho}_{m,n}(t)e^{\xi\hat{b}^{\dag}}
e^{-\xi^*\hat{b}}\}.
\label{eq:chimn}
\end{equation}
From Eq.~(\ref{eq:lindblad}) we obtain the equation for the density matrix
$\hat{\rho}_{m,n}$
\begin{equation}
\label{eq:unpmeas}
\frac{\partial}{\partial t}\hat{\rho}_{m,n}=-i\omega_{mn}
\hat{\rho}_{m,n}-i\lambda(\omega_m\hat{q}\hat{\rho}_{m,n}-
\hat{\rho}_{m,n}\hat{q}\omega_n)+L_D[\hat{\rho}_{m,n}],
\end{equation}
where
\begin{eqnarray}
\omega_n & = &\frac{E_n}{\hbar},\\
\omega_{mn} & = &\omega_m-\omega_n .
\end{eqnarray}
Equation (\ref{eq:unpmeas}) may be solved similarly as in
Sec.~\ref{sec:solution}.  When the detector is initially at equilibrium then
$\chi_{mn}(\xi ,\xi^*;0)=\chi_{mn}(0)\exp(-\xi\xi^*\bar{n}(T))$.  As in
Sec.~\ref{sec:solution} we take the characteristic function of the form
(\ref{eq:chi}) and obtain the equations for the coefficients $C_{j,k}$
\begin{eqnarray}
\frac{\partial C_{0,0}}{\partial t} & = & -i\omega_{mn}\left(
1+\lambda(C_{1,0}+C_{0,1})\right),\label{eq:coef1}\\
\frac{\partial C_{1,0}}{\partial t} & = & (i\Omega 
-\gamma_{\mathrm{eff}})C_{1,0}+i\lambda(\omega_n-\omega_{mn}C_{1,1}),\\
\frac{\partial C_{0,1}}{\partial t} & = & -(i\Omega 
+\gamma_{\mathrm{eff}})C_{0,1}-i\lambda(\omega_m+\omega_{mn}C_{1,1}),\\
\frac{\partial C_{1,1}}{\partial t} & = & (\gamma_{\uparrow}
-\gamma_{\downarrow})C_{1,1}+\gamma_{\uparrow}
\label{eq:coef4}
\end{eqnarray}
with the initial conditions $C_{0,0}(0)=0$, $C_{1,0}(0)=0$, $C_{0,1}(0)=0$,
$C_{1,1}(0)=\bar{n}(T)$. Here
\begin{equation}
\gamma_{\mathrm{eff}}=\frac{1}{2}(\gamma +\gamma_{\downarrow}
-\gamma_{\uparrow}).
\end{equation}
The solutions of Eqs.~(\ref{eq:coef1})--(\ref{eq:coef4}) are
\begin{eqnarray}
C_{1,1}(t) & = &\bar{n}(T),\label{eq:c11}\\
C_{1,0}(t) & = & i\lambda
\frac{\omega_n-\omega_{mn}\bar{n}(T)}{\gamma_{\mathrm{eff}}-i\Omega}
\left(1-e^{(i\Omega -\gamma_{\mathrm{eff}})t}\right),\\
C_{0,1}(t) & = & -i\lambda
\frac{\omega_m+\omega_{mn}\bar{n}(T)}{\gamma_{\mathrm{eff}}+i\Omega}
\left(1-e^{-(i\Omega +\gamma_{\mathrm{eff}})t}\right),\\
C_{0,0}(t) & = & -i\omega_{mn}t+\lambda^2\omega_{mn}
\frac{\omega_n-\omega_{mn}\bar{n}(T)}{\gamma_{\mathrm{eff}}-i\Omega}
\left(t+\frac{1}{\gamma_{\mathrm{eff}}-i\Omega}
\left(e^{(i\Omega -\gamma_{\mathrm{eff}})t}-1\right)\right)\nonumber\\
 &  & -\lambda^2\omega_{mn}
\frac{\omega_m+\omega_{mn}\bar{n}(T)}{\gamma_{\mathrm{eff}}+i\Omega}
\left(t+\frac{1}{\gamma_{\mathrm{eff}}+i\Omega}
\left(e^{-(i\Omega +\gamma_{\mathrm{eff}})t}-1\right)\right).
\label{eq:c00} 
\end{eqnarray}
Using Eqs.~(\ref{eq:chi}) and (\ref{eq:c00}) we find that the non-diagonal
elements of the density matrix of the measured system become small as the time
$t$ grows. This represents the decoherence induced by the measurement. The
diagonal elements of the density matrix do not change.

\section{\label{sec:pert-meas}Measurement of the perturbed system}

The operator $\hat{V}(t)$ represents the perturbation of the unperturbed
Hamiltonian $\hat{H}_0+\hat{H}_1$. We will take into account the influence of
the operator $\hat{V}$ by the perturbation method, assuming that the strength of
the interaction $\lambda$ between the system and detector is large.

The density matrix at time $t$ is related to the initial density matrix by the
equation $\hat{\rho}(t)=S(t)\hat{\rho}(0)$. The superoperator $S$ obeys the
equation
\begin{equation}
\label{eq:sup}
\frac{\partial}{\partial t}S=LS,
\end{equation}
where the Liouvillian superoperator $L$ is defined by the equation
\begin{equation}
L\hat{\rho}=\frac{1}{i\hbar}[\hat{H},\hat{\rho}]+L_D[\hat{\rho}].
\end{equation}
Here $\hat{H}$ and $L_D$ are defined by Eqs.~(\ref{eq:ham}) and (\ref{eq:LD}),
respectively.  We can write $L=L_0+L_V$, where $L_V$ is a small
perturbation, defined by the equation
\begin{equation}
L_V\hat{\rho}=\frac{1}{i\hbar}[\hat{V},\hat{\rho}].
\label{eq:LV}
\end{equation}
We expand the superoperator $S$ into powers of $V$
\[
S=S^{(0)}+S^{(1)}+S^{(2)}+\cdots .
\]
Then from Eq.~(\ref{eq:sup}) it follows
\begin{eqnarray}
\frac{\partial}{\partial t}S^{(0)} & = & L_0S^{(0)},\label{eq:sup0}\\
\frac{\partial}{\partial t}S^{(i)} & = & L_0S^{(i)}+L_VS^{(i-1)}.
\label{eq:supi} 
\end{eqnarray}
The formal solutions of Eqs.~(\ref{eq:sup0}) and (\ref{eq:supi}) are
\begin{equation}
S^{(0)}=e^{L_0t}
\end{equation}
and
\begin{equation}
S^{(i)}=\int^t_0dt_1S^{(0)}(t-t_1)L_VS^{(i-1)}(t_1).
\end{equation}
In the second-order approximation we have
\begin{eqnarray}
S(t) & = & S^{(0)}(t)+\int^t_0dt_1S^{(0)}(t-t_1)L_VS^{(0)}(t_1)\nonumber\\
& &+\int^t_0dt_1\int^{t_1}_0dt_2S^{(0)}(t-t_1)L_VS^{(0)}(t_1-t_2)
L_VS^{(0)}(t_2).
\label{eq:sup2} 
\end{eqnarray}

Let the initial density matrix of the system and detector is
\begin{equation}
\hat{\rho}(0)=|i\alpha\rangle\langle i\alpha |\otimes\hat{\rho}_D ,
\label{eq:rhoinit}
\end{equation}
where $\hat{\rho_D}$ is the density matrix of the detector.  The
probability of the jump from the level $|i\alpha\rangle$ to the level
$|f\alpha_1\rangle$ during the measurement is
\begin{equation}
W(i\alpha\rightarrow f\alpha_1,t)=\Tr\{|f\alpha_1\rangle\langle
f\alpha_1|\hat{\rho}(t)\} .
\label{eq:jump0}
\end{equation}
The unperturbed evolution does not change the energy of the measured system,
therefore, we can write as
\begin{equation}
S^{(0)}(t)\left[|m\alpha\rangle\langle n\alpha '|\otimes\hat{\rho}_D\right]
=|m\alpha\rangle\langle n\alpha '|\otimes S_{m\alpha ,n\alpha'}^{(0)}(t)
\hat{\rho}_D.
\label{eq:supdet}
\end{equation}
Equation (\ref{eq:supdet}) defines a new superoperator $S_{m\alpha
  ,n\alpha'}^{(0)}$ acting only on the density matrix of the detector.  The
indices $m\alpha$ and $n\alpha'$ in $S_{m\alpha ,n\alpha'}^{(0)}$ denote the
states of the measured system. From Eq.~(\ref{eq:supdet}) it follows that the
superoperator $S_{m\alpha ,m\alpha}^{(0)}$ with equal indices does not change
the trace of the density matrix $\hat{\rho}_D$, since the trace of the full
density matrix of the measured system and the detector must remain unchanged
during the evolution.

We assume that diagonal matrix elements of the perturbation operator $V$ are
zeros. Inserting the expression $\hat{\rho}(t)=S(t)\hat{\rho}(0)$ into
Eq.~(\ref{eq:jump0}) and using equation (\ref{eq:sup2}) for the superoperator
$S(t)$, we obtain the jump probability
\begin{eqnarray}
W(i\alpha\rightarrow f\alpha_1,t) & = &\frac{1}{\hbar^2}
\int^t_0dt_1\int^{t_1}_0dt_2
\Tr\left\{|f\alpha_1\rangle\langle f\alpha_1|\right.\nonumber \\
& & \times\left(S^{(0)}(t-t_1)\hat{V}S^{(0)}(t_1-t_2)
[S^{(0)}(t_2)\hat{\rho}(0)]\hat{V}\right.\nonumber\\
 &  & +S^{(0)}(t-t_1)[S^{(0)}(t_1-t_2)\hat{V}S^{(0)}(t_2)
\hat{\rho}(0)]\hat{V}\nonumber\\
& & -S^{(0)}(t-t_1)\hat{V}S^{(0)}(t_1-t_2)\hat{V}S^{(0)}(t_2)
\hat{\rho}(0)\nonumber \\
& & \left.\left.-S^{(0)}(t-t_1)[S^{(0)}(t_1-t_2)[S^{(0)}(t_2)\hat{\rho}(0)]
\hat{V}]\hat{V}\right)\right\}.
\label{eq:jumptmp1}
\end{eqnarray}
From Eqs.~(\ref{eq:supdet}) and (\ref{eq:rhoinit}) it follows that the last two
terms in Eq.~(\ref{eq:jumptmp1}) contain the scalar product $\langle
f\alpha_1|i\alpha\rangle$. Since the states $|f\alpha_1\rangle$ and
$|i\alpha\rangle$ are orthogonal, the last two terms in Eq.~(\ref{eq:jumptmp1})
are zeros.  Therefore, the jump probability is
\begin{eqnarray}
W(i\alpha\rightarrow f\alpha_1,t) & = &\frac{1}{\hbar^2}
\int^t_0dt_1\int^{t_1}_0dt_2
\Tr\left\{|f\alpha_1\rangle\langle f\alpha_1|\right.\nonumber \\
& & \times\left(S^{(0)}(t-t_1)\hat{V}S^{(0)}(t_1-t_2)
[S^{(0)}(t_2)\hat{\rho}(0)]\hat{V}\right.\nonumber\\
 &  & +\left.\left.S^{(0)}(t-t_1)[S^{(0)}(t_1-t_2)\hat{V}S^{(0)}(t_2)
\hat{\rho}(0)]\hat{V}\right)\right\}.
\label{eq:jump1}
\end{eqnarray}
From Eq.~(\ref{eq:jump1}), using expression for the initial density matrix of
the system and the detector (\ref{eq:rhoinit}) and equation (\ref{eq:supdet}),
we have
\begin{eqnarray}
W(i\alpha\rightarrow f\alpha_1,t)&=&\frac{1}{\hbar^2}
|V_{i\alpha ,f\alpha_1}|^2\int^t_0dt_1\int^{t_1}_0dt_2\Tr\left\{
S_{f\alpha_1 ,f\alpha_1}^{(0)}(t-t_1)\right.\nonumber \\
& & \left.\times\left(S_{i\alpha ,f\alpha_1}^{(0)}(t_1-t_2)
+S_{f\alpha_1 ,i\alpha}^{(0)}(t_1-t_2)\right)
S_{i\alpha ,i\alpha}^{(0)}(t_2)\hat{\rho}_D\right\}.
\label{eq:jumptmp2} 
\end{eqnarray}
The superoperator $S_{f\alpha_1 ,f\alpha_1}^{(0)}(t-t_1)$ preserves the trace of
the detector's density matrix, therefore, the jump probability equals to
\begin{eqnarray}
W(i\alpha\rightarrow f\alpha_1,t)&=&\frac{1}{\hbar^2}
|V_{i\alpha ,f\alpha_1}|^2\int^t_0dt_1\int^{t_1}_0dt_2\Tr\left\{\left(
S_{i\alpha ,f\alpha_1}^{(0)}(t_1-t_2)\right.\right.\nonumber \\
&&+\left.\left. S_{f\alpha_1 ,i\alpha}^{(0)}(t_1-t_2)\right)
S_{i\alpha ,i\alpha}^{(0)}(t_2)\hat{\rho}_D\right\}.
\label{eq:jump2} 
\end{eqnarray}
Defining a new characteristic function similarly as in Eq.~(\ref{eq:chimn})
\begin{equation}
\chi_{i\alpha ,f\alpha_1}(\xi ,\xi^*;t_1,t_2)=\Tr\{
e^{\xi\hat{b}^{\dag}}e^{-\xi^*\hat{b}}
S_{i\alpha ,f\alpha_1}^{(0)}(t_1-t_2)S_{i\alpha, i\alpha}^{(0)}(t_2)\hat{\rho}_D
\}
\label{eq:chi2}
\end{equation}
the jump probability (\ref{eq:jump2}) can be expressed as
\begin{equation}
\label{eq:jumpprob}
W(i\alpha\rightarrow f\alpha_1,t)=\frac{1}{\hbar^2}|V_{i\alpha ,f\alpha_1}|^2
\int^t_0dt_1\int^{t_1}_0dt_2\left(\chi_{i\alpha ,f\alpha_1}(0,0;t_1,t_2)
+\chi_{f\alpha_1,i\alpha}(0,0;t_1,t_2)\right). 
\end{equation}

The detector initially (at $t=0$) is in the thermal equilibrium with the
temperature $T$, $\hat{\rho}_D=\hat{\rho}_T$ (Eq.~(\ref{eq:rhoT})). The initial
characteristic function is $\chi_{i\alpha ,f\alpha_1}(\xi
,\xi^*;0,0)=\exp\left(-\xi\xi^*\bar{n}(T) \right)$. Using the results of the
Sec.~\ref{sec:unpert-meas} (Eqs.~(\ref{eq:chi}) and
(\ref{eq:c11})--(\ref{eq:c00}) ), we obtain the characteristic function of the
density matrix at time $t_2$
\begin{eqnarray}
\chi_{i\alpha ,f\alpha_1}(\xi ,\xi^*;t_2,t_2)&=&\Tr\{e^{\xi\hat{b}^{\dag}}
e^{-\xi^*\hat{b}}S_{i\alpha ,i\alpha}^{(0)}(t_2)\hat{\rho}_D\} \nonumber\\
&=&\exp\left(\frac{i\xi\lambda\omega_i}{\gamma_{\mathrm{eff}}
-i\Omega}\left(1-e^{i(\Omega -\gamma_{\mathrm{eff}})t_2}\right)
\right) \nonumber \\
&&\times\exp\left(\frac{i\xi^*\lambda\omega_i}{\gamma_{\mathrm{eff}}
+i\Omega}\left(1-e^{-i(\Omega
+\gamma_{\mathrm{eff}})t_2}\right)-\xi\xi^*\bar{n}(T)\right).
\label{eq:chit1}
\end{eqnarray}
Taking the function $\chi_{i\alpha ,f\alpha_1}(\xi ,\xi^*;t_2,t_2)$ from
Eq.~(\ref{eq:chit1}) as the initial characteristic function and proceding
further as in Sec.~\ref{sec:unpert-meas}, we have the value of the
characteristic function, defined by Eq.~(\ref{eq:chi2}) with the parameters
$\xi=\xi^{*}=0$
\begin{eqnarray}
\chi_{i\alpha ,f\alpha_1}(0,0;t_1,t_2) & = &\exp\bigg(
-i\omega_{i\alpha ,f\alpha_1}(t_1-t_2) \nonumber \\
 &  & +\lambda^2\omega_{if}
\frac{\omega_f-\omega_{if}\bar{n}(T)}{\gamma_{\mathrm{eff}}-i\Omega}
\left(t_1-t_2+\frac{1}{\gamma_{\mathrm{eff}}-i\Omega}
\left(e^{(i\Omega -\gamma_{\mathrm{eff}})(t_1-t_2)}-1\right)\right) \nonumber \\
 &  & -\lambda^2\omega_{if}
\frac{\omega_i+\omega_{if}\bar{n}(T)}{\gamma_{\mathrm{eff}}+i\Omega}
\left(t_1-t_2+\frac{1}{\gamma_{\mathrm{eff}}+i\Omega}
\left(e^{-(i\Omega +\gamma_{\mathrm{eff}})(t_1-t_2)}-1\right)\right) \nonumber \\
 &  & +\frac{\lambda^2\omega_{if}\omega_i}{(\gamma_{\mathrm{eff}}-i\Omega)^2}
\left(1-e^{(i\Omega -\gamma_{\mathrm{eff}})t_2}\right)
\left(1-e^{(i\Omega -\gamma_{\mathrm{eff}})(t_1-t_2)}\right) \nonumber \\
 &  & -\frac{\lambda^2\omega_{if}\omega_i}{(\gamma_{\mathrm{eff}}
+i\Omega)^2}\left(1-e^{-(i\Omega +\gamma_{\mathrm{eff}})t_2}\right)
\left(1-e^{-(i\Omega +\gamma_{\mathrm{eff}})(t_1-t_2)}\right)\bigg).
\end{eqnarray}
Here
\begin{equation}
  \omega_{i\alpha ,f\alpha_1}=\omega_{if}+
  \frac{1}{\hbar}\left(E_1(i,\alpha)-E_1(f,\alpha_1)\right).
\end{equation}

\subsection{Approximations}

When the dissipation is fast, i.e., the dissipation time is much less than the
period of the oscillator, we have $\Omega\ll\gamma_{\mathrm{eff}}$.  Then
\begin{eqnarray*}
\chi_{i\alpha ,f\alpha_1}(0,0;t_1,t_2)&=&\exp\left(-i\omega_{i\alpha ,f\alpha_1}
(t_1-t_2)\right)\\
&&\times\exp\left(-(1+2\bar{n}(T))\frac{\lambda^2\omega^2_{if}}{\gamma_{\mathrm{eff}}}
\left(t_1-t_2+\frac{1}{\gamma_{\mathrm{eff}}}
\left(e^{-\gamma_{\mathrm{eff}}(t_1-t_2)}-1\right)\right)\right).
\end{eqnarray*}
The probability of the jump from the level $|i\alpha\rangle$ to the level
$|f\alpha_1\rangle$ during the measurement according to
Eq.~(\ref{eq:jumpprob}) is
\begin{eqnarray}
W(i\alpha\rightarrow f\alpha_1,t) & = &\frac{2t}{\hbar^2}
|V_{i\alpha ,f\alpha_1}|^2\re\int^t_0du
\left(1-\frac{u}{t}\right)e^{i\omega_{f\alpha_1,i\alpha}u}\nonumber\\
 &  &\times\exp\left(
-\frac{(1+2\bar{n}(T))\lambda^2\omega^2_{if}}{\gamma_{\mathrm{eff}}}
\left(u+\frac{1}{\gamma_{\mathrm{eff}}}
(e^{-\gamma_{\mathrm{eff}}u}-1)\right)\right).
\label{eq:41}
\end{eqnarray}
We introduce the function
\begin{equation}
\Phi(t)_{f\alpha_1,i\alpha}=|V_{i\alpha ,f\alpha_1}|^2\exp\left(
\frac{i}{\hbar}[E_1(f,\alpha_1)-E_1(i,\alpha)]t\right) 
\end{equation}
and the Fourier transformation of $\Phi(t)_{f\alpha_1,i\alpha}$
\begin{equation}
G(\omega)_{f\alpha_1,i\alpha}=\frac{1}{2\pi}
\int^{\infty}_{-\infty}dt\Phi(t)_{f\alpha_1,i\alpha}\exp(-i\omega t).
\end{equation}
Then we can rewrite Eq.~(\ref{eq:41}) in the form
\begin{equation}
\label{eq:result}
W(i\alpha\rightarrow f\alpha_1,t)=\frac{2\pi t}{\hbar^2}
\int^{\infty}_{-\infty}d\omega G(\omega)_{f\alpha_1,i\alpha}P(\omega)_{if},
\end{equation}
where
\begin{eqnarray}
P(\omega)_{if} & = &\frac{1}{\pi}\re\int^t_0du\left(
1-\frac{u}{t}\right)\exp\left(i(\omega -\omega_{if})u\right)\nonumber\\
 &  &\times\exp\left(
-\frac{(1+2\bar{n}(T))\lambda^2\omega^2_{if}}{\gamma_{\mathrm{eff}}}
\left(u+\frac{1}{\gamma_{\mathrm{eff}}}
(e^{-\gamma_{\mathrm{eff}}u}-1)\right)\right).
\end{eqnarray}
The equation (\ref{eq:result}) is of the form, obtained by Kofman and Kurizki
\cite{Kofman}, assuming the ideal instantaneous projections.  The function
$P(\omega)_{if}$ is the measurement-modified shape of the spectral line
(Refs.~\cite{rus2,Kofman,Kofman2}). Here we have shown that
Eq.~(\ref{eq:result}) can be derived from more realistic model as well. The
assumption that dissipation is fast, $\Omega\ll\gamma_{\mathrm{eff}}$ is
crucial. Without this assumption the jump probability cannot have the form of
Eq.~(\ref{eq:result}), since then $\chi_{i\alpha ,f\alpha_1}(0,0;t_1,t_2)$
depends not only on the difference $t_1-t_2$ but also on $t_2$.

When $\lambda$ is big then to the integral in Eq.~(\ref{eq:41}) contribute only
small values of $u$ and we can expand the exponent
$\exp(-\gamma_{\mathrm{eff}}u)$ into Taylor series keeping the first three terms
only. We obtain the jump rate
\[
R(i\alpha\rightarrow f\alpha_1)\approx\frac{2}{\hbar^2}
|V_{i\alpha ,f\alpha_1}|^2\re\int^{\infty}_0du\,\exp\left(
i\omega_{f\alpha_1,i\alpha}u-\frac{1}{2}(1+2\bar{n}(T))
\lambda^2\omega^2_{if}u^2\right)
\]
or
\begin{equation}
R(i\alpha\rightarrow f\alpha_1)\approx
\frac{2|V_{i\alpha ,f\alpha_1}|^2}{\hbar^2\lambda |\omega_{if}|}
\sqrt{\frac{\pi}{2(1+2\bar{n}(T))}}
\end{equation}
The obtained decay rate inversely proportional to the measurement strength
$\lambda$. The measurement strength appears in the equations multiplied by
$\sqrt{1+2\bar{n}(T)}$, therefore, the effect of the measurement increases as
the temperature of the detector grows.

\section{\label{sec:concl}Conclusions}

We analyze the quantum Zeno effect using the irreversible model of the
measurement. The detector is modeled as a harmonic oscillator, initially being
at the thermal equilibrium. The interaction of the detector with the system is
modeled similarly as in Ref.~\cite{rus2}.  The Lindblad-type master equation for
the detectors density matrix is solved analytically. An equation for the
probability of the jump between measured system's states during the measurement,
similar to that of Refs.~\cite{rus2,Kofman,Kofman2}, is obtained
(\ref{eq:result}).  From the used model it follows that the increase of the
detector's temperature leads to the enhancement of the quantum Zeno or quantum
anti-Zeno effects.

\begin{acknowledgments}
  I wish to thank Professor B. Kaulakys for his suggestion of the problem, for
  encouragement, stimulating discussions, and critical remarks.
\end{acknowledgments}


\end{document}